# Defect model for the mixed mobile ion effect

Vladimir Belostotsky

1005 Curtis Place, Rockville, MD 20852, USA

E-mail: vladbel@erols.com

This paper presents a new defect model for the mixed mobile ion effect. The essential physical concept involved is that simultaneous migration of two unlike mobile ions in mixed ionic glass is accompanied by expansion or contraction of the guest-occupied sites with distortion of surrounding glass matrix; in many cases, an intensity of the local stresses in glass matrix surrounding ionic sites occupied by foreign ions is much greater than, or at least comparable to the glass network binding energy. Hence, when the stress exceeds the breaking threshold, relaxation occurs almost immediately via the rupture of the bonds in the nearest glass matrix with generation of pairs of intrinsic structural defects. The specificity of the mechanism of defect generation leads to the clustering of negatively charged defects, so that rearranged sites act as high energy anion traps in glass matrix. This results in the immobilization of almost all minority mobile species and part of majority mobile species, so mixed mobile ion glass behaves as single mobile ion glass of much lower concentration of charge carriers. Generation of defects leads also to the depolymerization of glass network, which in turn results in the reduction of the glass viscosity and $T_g$ as well as in the compaction of glass structure (thermometer effect). In the spectra of mechanical losses of mixed alkali glasses it reflects as a shift of the maximum in mechanical losses corresponding to the glass transition to lower temperatures, and the dramatic increase of the maximum corresponding to the movement of non-bridging oxygens (so-called mixed alkali peak). The magnitude of the mixed mobile ion effect is defined by the size mismatch of unlike mobile ions, their total and relative concentrations, the binding energy of the glass-forming network, and temperature. Although the proposed model is based upon the exploration of alkali silicate glass-forming system, the approach developed here can be easily adopted to other mixed ionic systems such as crystalline and even liquid ionic conductors.

## 1. INTRODUCTION

In oxide glasses, where two dissimilar alkali species coexist, many physical properties, most notably ionic conductivity and viscosity, demonstrate remarkable deviation from the linearity when one alkali cation substitutes another at total fixed concentration of alkalis [1, 2, 3]. The same effect occurs in related chalcogenide systems, in glasses containing protons, $Tl^+$, $Cu^+$, and $Ag^+$ ions, in glasses subjected to ion-exchange [4], and even in crystalline solids [5] and liquids [3, 6]. Similar effects are observed in mixed alkaline-earth glasses [7] and mixed anion systems [8]. Thus, the problem appears to be fairly general, i.e., its origin is not related only to alkali cations or it should not lie in the non-crystalline structure.

This phenomenon is usually referred to as mixed alkali effect, mixed cation effect, or mixed mobile ion effect (MMIE), and although it is known since 1883 (initially as the 'thermometer' effect), there is no general agreement concerning its origin.

Various models were proposed to account for the MMIE in ionic mobility [6, 9, 10, 11, 12, 13, 14, 15, 16] and viscosity [17, 18, 19], however, to the best of the author's knowledge, up to date no model has been proposed to agree, at least qualitatively, with all experimental facts. What is common to all the models, though, is an implicit postulation that ionic movement only occurs between so-called regular ionic sites, in oxide glasses they are comprised of oxygens of which one is non-bridging oxygen and the rest are bridging, and that the local mechanical stresses created by small ions entering large sites and large ions entering small sites are absorbed somehow by the surrounding network-forming matrix, so that the structure of the matrix itself remains largely unaffected when a mobile ion enters a foreign site. The approach dominating the current literature on the MMIE considers the ionic transport in mixed ionic systems as 'site preferred': the unlike ions are unwilling to visit each other's sites because of a mismatch between the requirements of the mobile ion with what the foreign site and the doorway to it could offer in terms of a cavity size and a number of the nearest neighbors [20, 21].

The idea that the MMIE is related somehow to the defects is not new. LaCourse [17] has proposed a defect model for mixed alkali effect where alkali cations in foreign sites were defined as 'mixed alkali defects'. According to that model, mechanical and electrical strains localized on defect sites are causing strains in, and weakening of the network former – bridging oxygen bonds. Mixed alkali defects thus reduce the viscosity and account for the observed minimum with compositional variations. In the transition region, stress and structural relaxation causes mixed alkali defects to convert to normal sites with time. Diffusion/conductivity behavior in the mixed alkali defect model is predicted by assuming to be site preferred; alkali cations may diffuse along the foreign sites, though, but at reduced rate.

The attempts to examine the role of 'true' intrinsic structural defects in the network-forming matrix in the MMIE also have been made [22, 23, 24, 25]. Hayward [22] concluded that the presence of non-bridging oxygen anions is responsible, directly or indirectly, for the MMIE in oxide glasses, however actual relationship was not established.

In the present paper a 'true' defect model for the MMIE is proposed. It is based upon the premise that the MMIE is caused by the intrinsic structural defects, in oxide glasses these are non-bridging oxygen anions, generated in the network-forming matrix, which in turn result from the expansion or contraction of the sites when mobile ions of different size enter foreign sites. It will be demonstrated that this approach offers a natural, straightforward, and consistent explanation of the MMIE in all its features.

## 2. DISSIMILAR ION MOVEMENT AND NETWORK-FORMING MATRIX RESPONSE

The contemporary model for ionic transport in glass holds that below the transformation temperature, ionic species are strongly associated with their local environment, which is defined as regular ionic sites. Like in ionic crystals, the displacement of the ions out from their equilibrium sites is viewed as a thermally activated process of generating interstitial ions and open and available sites [26]. Thus, the ion movement in disordered structures, at least in the dc regime and in the ac regime at not too high frequencies, can be viewed as barrier hopping between vacant sites in a random energy landscape. According to the percolation theory, ion conduction is determined by the infinite cluster constructed from the ionic sites which provides a connected pathway for the carriers through the whole system with the easiest hopping transitions [20].

It is also widely accepted that cations in glass are primarily responsible for the determining their local environment. Each cation tends to be surrounded regularly and densely by anions at distances depending on the respective cation size [27]. For example, $Li^+$ shows a strong tendency to fourfold coordination with oxygen atoms, whereas for $Na^+$ the most frequent coordination with oxygen is five or six [28], for $K^+$ the preferred coordination number is six to eight, and for $Cs^+$ it is eight to twelve [29]. Ionic radii of alkali cations and their coordination numbers are summarized in the Table I.

Experimental investigations show that specific local environment is clearly identifiable for each alkali species both in single and mixed alkali-silicate glasses [30, 31]. This finding has two possible explanations: First, this may occur if the cation migration is the site preferred. This means that in mixed alkali glass cations can jump only to the sites previously occupied by the cations of the same type. This notion is difficult to accept: if it were the case, ion exchange diffusion could not be possible. The other possibility is that the rearrangement in the nearest ion environment accompanying cation movement is fast enough to accommodate the changes. Our further consideration is based on the latter premise, namely that site rearrangement processes occur on the time scale close to that of the ion migration.

Our starting point is initially defect-free $SiO_2$ glass matrix modified by two unlike alkali species, $A^+$ and $R^+$, where all modifiers reside in regular sites.

Since there are no preferable vacant ionic sites for both species, cations can randomly jump occupying both host and foreign regular vacant sites. However, in mixed alkali glass a cation occupying a foreign site must be considered as a defect, which induces local strain fields in the glass matrix surrounding an ionic site.

Indeed, when $A^+$-ion moves into an $R$-site, it tends to rearrange the site according to its steric requirements. If the $A^+$-ion is larger, a mismatch in the sizes of *[AO$_K$]* and *[RO$_N$]* cation-centered polyhedra induces compressive stress in its local environment in radial direction and tensile stress in the coordination shells around *[AO$_K$]* polyhedron [32]. The total strain energy, $E_S$, associated with the substitution of the spherical polyhedron with radius $r_R$ for the polyhedron with radius $r_A$ in homogeneous medium, is given by the Frenkel's equation [33, p.11]:

$$E_S = 8\pi G r_R (r_A - r_R)^2 \qquad (1)$$

where $G$ is the shear modulus. The shape of the polyhedra is taken to be spherical for calculation purposes. Values calculated for $E_S$ for various guest - host pairs of alkali cations in silicate glasses are summarized in Table II. Calculations were made based on the radii of cation-centered coordination polyhedra given in the Table I, and in assumption that the polyhedra are largely incompressible. The shear modulus was taken $G \approx 3.05 \times 10^{10}$ N/m$^2$ [34]. For the comparison, the values for the total strain energy obtained by *ab initio* molecular orbital calculations for a sodium ion in the site previously occupied by lithium and for a potassium ion in the former lithium site are *1.46 eV*, and *6.95 eV* respectively [35].

TABLE I. Cation coordination numbers (*CN*), ionic radii ($r_i$), cation – oxygen internuclear distances ($r_{i-O}$) [119, 120, 121] and calculated radii of coordination polyhedra ($r_{[i-O]}$), radii ($r_{FCS}$) and circumference lengths of the first coordination shells ($L_{FCS}$) around coordination polyhedra in silicates.

| Ion | CN | $r_i$, pm | $r_{i-O}$, pm | $r_{[i-O]}$, pm | $r_{FCS}$, pm | $L_{FCS}$, pm |
|---|---|---|---|---|---|---|
| Li | 4 | 59 | 196 | 333 | 470 | 2951 |
|    | 6 | 76 | 213 | 350 | 487 | 3058 |
| Na | 5 | 100 | 237 | 374 | 511 | 3209 |
|    | 6 | 102 | 239 | 376 | 513 | 3222 |
|    | 7 | 112 | 249 | 386 | 523 | 3284 |
|    | 8 | 118 | 255 | 392 | 529 | 3322 |
|    | 9 | 124 | 261 | 398 | 535 | 3360 |
| K  | 6 | 138 | 275 | 412 | 549 | 3448 |
|    | 7 | 146 | 283 | 420 | 557 | 3498 |
|    | 8 | 151 | 288 | 425 | 562 | 3529 |
|    | 9 | 155 | 292 | 429 | 566 | 3554 |
|    | 10 | 159 | 296 | 433 | 570 | 3580 |
|    | 12 | 164 | 301 | 438 | 575 | 3611 |
| Cs | 8 | 174 | 311 | 448 | 585 | 3674 |
|    | 9 | 178 | 315 | 452 | 589 | 3699 |
|    | 10 | 181 | 318 | 455 | 592 | 3718 |
|    | 11 | 185 | 322 | 459 | 596 | 3743 |
|    | 12 | 188 | 325 | 462 | 599 | 3762 |
| Rb | 8 | 161 | 298 | 435 | 572 | 3592 |
|    | 9 | 163 | 300 | 437 | 574 | 3605 |
|    | 10 | 166 | 303 | 440 | 577 | 3624 |
|    | 11 | 169 | 306 | 443 | 580 | 3642 |
|    | 12 | 172 | 309 | 446 | 583 | 3661 |
|    | 14 | 183 | 320 | 457 | 594 | 3730 |
| O  |   | 137 |   |   |   |   |

TABLE II. Strain energy, $E_S$, strain intensity index, $N_d$, difference in cation coordination number ($\Delta CN$), and difference in length of the circumference of the first coordination shell ($\Delta L_{FCS}$) for various alkali cation pairs in mixed alkali silicate glasses

| Ionic pairs | $E_S$, eV | $N_d$ | $\Delta CN$ | $\Delta L_{FCS}$, pm |
|---|---|---|---|---|
| Li-Na | 2.7 - 4.5 | 1 – 3 | 1 - 2 | 271 |
| Li-K | 13.5 – 15.9 | 3 - 8 | 4 – 6 | 578 |
| Li-Cs | 23.7 – 26.5 | 6 - 14 | 6 - 8 | 767 |
| Li-Rb | 18.2 – 24.5 | 5 - 12 | 6 - 8 | 673 |
| Na-K | 4.3 – 5.9 | 1 - 3 | 2 - 4 | 307 |
| Na-Cs | 11.2 – 13.3 | 3 - 7 | 4 - 6 | 496 |
| Na-Rb | 7.4 – 11.8 | 2 - 6 | 4 - 6 | 402 |
| K-Cs | 1.8 – 2.8 | 1 - 2 | 2 - 4 | 189 |
| K-Rb | 0.5 – 2.1 | 0 - 1 | 2 - 4 | 95 |
| Cs-Rb | 0.5 | 0 - 1 | 0 - 2 | |

Local stress field decays as the inverse square root of the distance from the defect [32]. Therefore, for many guest - host pairs high-stress region is not localized on the cation-centered polyhedron but may involve next coordination shells. The size of the high-stress region depends on the energy of the defect, finally on the mismatch in ionic radii of the host and foreign ions.

To characterize the intensity of the strain we specify a strain intensity index

$$N_d = E_S/E_{Me-O} \qquad (2)$$

where $E_{Me-O}$ is the chemical strength of a single network-forming *Me-O* bond. Using literature data on the strength of *Si-O* bonds in silicates which according to different authors varies from *2 eV* [36] to *4 eV* [37, 38, 39], the integer strain intensity index has been calculated for various ionic pairs and the results are summarized in the Table II.

As can be seen, in many cases $N_d$ is close to, or larger than unity indicating the fact that the intensity of the strain is greater than, or at least comparable to network binding energy. Therefore, it is logical to suggest that at least partial relaxation of the local stress applied to the surrounding glass matrix occurs almost immediately via either mechanical or/and thermally activated breaking of the stretched bonds in the nearest glass network, in silicate glasses these are predominantly the bridging *Si-O* bonds surrounding *[AO_K]* polyhedra, with generation of pairs of structural defects, oxygen vacancy centers and non-bridging oxygen anions [36].

Actually, this notion seems highly speculative and even heretical. It contradicts the postulates of the theory of solid state, specifically theory of defects is solids, according to which the compression should vanish because the volume of the whole body must increase by the amount equal to the difference between the volumes of the rigid sphere and that of the spherical cavity, while its density is not altered [32, p.184; 33, p.10]. It also contradicts the generally accepted viewpoint among glass scientists that due to the presence of residual looseness in glassy phase, a glass system should exhibit an anelastic response to mechanical stress where the local displacement of atomic particles at loose spots shall serve to relax the applied stress in the jammed structure in a reversible way [40, 41, 42, 43, 44, 45].

However, literature data on infrared (IR) spectra of ion-exchanged glasses offer lines of direct evidence that the larger-for-smaller ion replacement is accompanied by the formation of structural defects. As is well established, the IR spectra of silicate glasses are similar to those of silica glass, and IR reflectances are proportional to the concentration of mechanisms causing them [46]. There are several fundamental vibration bands for silica structure in the wavenumber range of 1200 - 450 cm$^{-1}$ observable by IR spectroscopy [47, 48]: the band near ~ 1050 cm$^{-1}$ is attributed to *Si-O* asymmetric stretching mode in the *Si-O-Si* bridging bonds, and the band with the maximum near ~950 cm$^{-1}$ is assigned to *Si-O* stretching mode of non-bridging oxygen (NBO). Lee *et al.* [49] studied IR reflection spectra of soda-lime silicate glass exposed to $KNO_3$ melt and reported a decrease in the peak amplitude of the band near ~ 1050 cm$^{-1}$, and a dramatic increase in the peak amplitude of the band near ~ 950 cm$^{-1}$ which is poorly resolved as a shoulder in the integrated reflection in the 800 – 1200 cm$^{-1}$ frequency range in the spectrum of the virgin glass (Fig. 1). Similar changes in IR reflection spectra of $Ag^+$-for-$Na^+$ ion-exchanged sodium silicate glass were observed by Yamane *et al.* [50]. Although currently accepted viewpoint holds that ion exchange does not involve changes in the structure of the glass-forming network [51, 52], those experimental observations clearly indicate that the larger-for-smaller ion interdiffusion is accompanied by the depolymerization of the glass network via the breaking of *Si-O-Si* bridging bonds and generation of structural defects, non-bridging oxygen anions and oxygen vacancies. Interestingly, a remarkable similarity in phenomenology related to the glass-network damage is observed in IR reflection spectra of silica glass irradiated by $Ar^+$ ions [53].

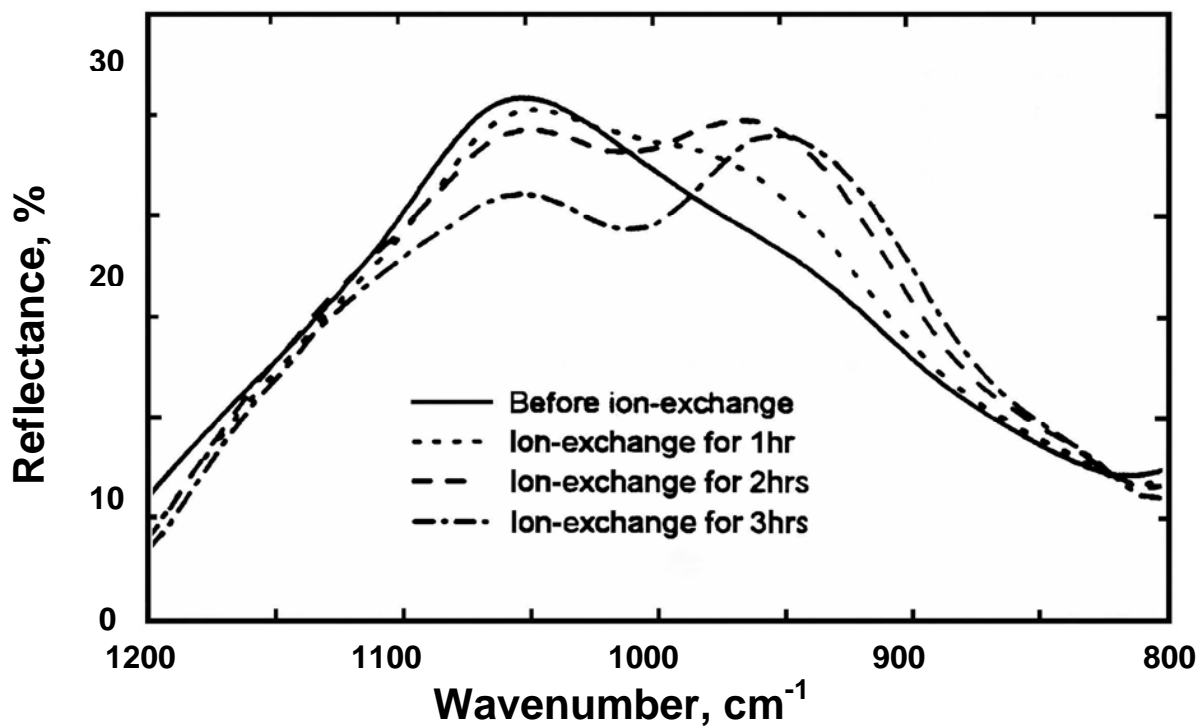

Fig. 1. Infrared reflection spectra of soda lime silicate glass ion-exchanged in molten $KNO_3$ for various times (Redrawn from Ref. 49)

Accordingly, when a smaller $R^+$-ion moves into a larger *A*-site, it also tends to rearrange the site in accordance with its steric requirements. Evidently, a former *A*-site occupied by $R^+$-ion shall shrink causing a corresponding distortion of the surrounding glass network with generation of structural defects. Indeed, in condensed materials, thermal pressure, $P_T$, caused by thermal movement of atomic particles is completely compensated by internal pressure, $P_i$, caused by the forces of attraction between particles [54]:

$$P_T = -\left(\frac{\partial F}{\partial V}\right)_T \tag{3}$$

$$P_i = \frac{dU}{dV} \tag{4}$$

$$P_T = P_i \tag{5}$$

where all symbols have their usual meaning. In solids, internal pressure coincides approximately with the microhardness, and can be calculated by the expression [54]:

$$P_i \cong \frac{(1-2\nu)H}{6(1+\nu)} \tag{6}$$

where $\nu$ is Poisson's ratio and $H$ is the modulus of elasticity. For silicate glasses $P_i$ = *4.0 - 6.5 GPa*. Total strain energy estimated by Uchino *et al.* [35] based on *ab initio* molecular orbital calculations for $Li^+$ in the former *Na* and *K* sites is *0.69* and *1.88 eV*, respectively, which is much less than that obtained for the reverse cases but it might be sufficient to cause generation of defects in the vicinity of distorted sites.

Again, literature data on ion-exchanged glasses bear out that when a smaller ion jumps into a larger site, the shrinkage of the site and generation of structural defects takes place. This must be apparent from the experiments on exposing a soda-lime glass to a melt of a lithium salt at a temperature where ion-exchange diffusion can occur. According to observations of Ernsberger [55] and Kistler [56], dipping a strip of glass into a lithium nitrate melt for only a few minutes produces a dense mat white surface whose microscopic examination shows to be torn into innumerable cracks. The cracks move by breaking individual bonds between atoms, and therefore can be regarded as a macroscopic probe for the bond breaking on the atomic level.

Our picture of structural rearrangement in mixed alkali glass wouldn't be complete if we do not mention the situation where the local stress fields of compression and tension in glass matrix are developed in concert with each other in an interacting manner when the pairs of unlike alkali cations are swapped around the neighboring sites. Although this point requires additional investigation, we suggest here that from the standpoint of the formation of defects, such coupled expansions and contractions shall be viewed as cumulative and not self-canceling.

Yet again, literature data on optical properties of mixed alkali glasses support the notion of defect formation in glass matrix caused by simultaneous movement of two unlike mobile ions: Agarwal and co-workers [57] have reported the existence of a minimum in optical band gap at equimolar concentrations of alkali cations in lithium – potassium borate glass, and attributed it to the formation of large number of NBOs in comparison to dilute foreign alkali regions. Kamitsos *et al.* [58, 59] have studied IR reflectance of mixed alkali borate glasses, and found that alkali mixing causes the partial destruction of $BØ_4^-$ groups in favor of their $BØ_2O^-$ isomeric triangles (here $Ø$ and $O^-$ denote bridging and non-bridging oxygens respectively), and that the fraction of non-bridging oxygens should exhibit a positive departure from linearity.

Thus, based upon the foregoing consideration one may conclude that although some additional experimental investigation is needed, a suggestion that in mixed mobile ion glasses simultaneous migration of unlike ions of different size causes a formation of structural defects in the immediate vicinity of the ionic sites where a host ion is substituted by a foreign one is quite plausible. Evidently, we shall expect that the number of generated defect pairs per ionic site is determined by the size mismatch of unlike ions and the binding energy of the glass-forming network. At temperatures below the glass transformation range, such a structural configuration is practically arrested because of long relaxation times of the network-forming matrix.

It is important to emphasize over again that the ion size mismatch is crucial in the MMIE. When the ionic radii of dissimilar ions are close to each other, the MMIE shall degenerate in all its features as it presumably occurs in *CuI-AgI-As$_2$Se$_3$* glasses [60].

One of the direct consequences of the bond breaking and generation of defects is the compaction of glass structure which is macroscopically observed as the 'thermometer effect' [61].

## 3. DEFECT GENERATION AND REARRANGEMENT OF IONIC SITES

As was pointed out in the previous section, in mixed alkali glasses structural defects arise from, and together with, rearrangement of ionic sites. Moreover, when the strain intensity index is appreciably larger than unity, site rearrangement can only occur through the defect formation in its vicinity, and it is thought to happen in a manner similar to the microcrack-extension (or even nanocrack-extension) mechanism rather than random formation of isolated defect pairs.

The processes by which elastic energy is converted into broken bonds on nanometric scale occur in the fundamentally discrete and not continuous medium, so the resistance to nanocrack propagation shall be characterized by the forces required to separate network-forming bonds successively. It is intuitively obvious that the preferred direction in which a nanocrack will take the path is associated with the lowest energy barrier, in silicate glass it is the weakest Si-O bond in the first coordination shell around an ionic site. The bond will snap when its total extension exceeds a bond-breaking distance. Calculation of the maximum elongation of the circumference of the first coordination shell around a cation-centered polyhedron for various alkali pairs shows that for the *Na-K* pair it exceeds tree angstroms, and for the *Li-Cs* pair it reaches almost eight angstroms, which are well beyond a bond-breaking threshold for a *Si-O* bond (see Table II).

Then, if the successful separation of the first atomic pair does not relieve the stress, the initiated nanocrack will be advancing from one bond to the other causing a cascade of bond breaking until the tensile stress is totally relieved, or until a nanocrack is arrested by so-called 'lattice-trapping' mechanism [62]. The trapping barrier can be overcome, though, by the thermal activation [63]. The time scale on which a nanocrack snaps an atomic bond and the associated relaxation is of the order of atomic vibrational period (~ $10^{-12}$ s) [63], which is close to the time scale of the ion migration process.

Evidently, the strain energy dissipated through nanocrack propagation is approximately a sum of the interaction energies of the separated atomic pairs [62]. Therefore, if we assume here that all strain energy dissipates trough the bond breaking, then the total number of bonds broken as a result of the site rearrangement is nothing more than the strain intensity index, $N_d$, introduced in the previous section and calculated from Eqn. (2). Thus, the strain intensity index has a physical meaning of the maximum possible number of defect pairs arising from the site rearrangement when a larger alkali cation enters a smaller foreign site.

A larger ion has a greater steric demand. Therefore, the site rearrangement shall involve also an incorporation of additional oxygens into the cation-centered polyhedron. Apparently, the vacant positions will be taken by non-bridging oxygens generated nearby due to their strong long-range electrostatic interaction with alkali cation.

A physical picture of the site rearrangement and generation on defects in the network-forming matrix when a smaller mobile ion jumps into a larger foreign site is less apparent. We feel there are a number of questions that are difficult to answer right away. Further insight is needed to clarify this point. Nevertheless, we presume here that generation of the defects accompanies the site rearrangement in both cases, larger-for-smaller and smaller-for-larger ion replacement.

We shall not be concerned now with the fate of the complimentary pairs of non-bridging oxygens, oxygen vacancies. It will be discussed further in the Section 5. However, it must be emphasized that although the charge neutrality must be maintained, it can only be maintained on average. This means that the sum over the net charge of all charged point defects must be zero, but it is erroneous to require that the charges of the opposite sign cancel each other on the microscopic scale. This is proper only for large volumes.

## 4. ANION TRAPS AND IONIC TRANSPORT IN MIXED MOBILE ION GLASSES

It is well established that atomic transport in solids takes place because of the presence of defects, and atomic diffusion can be accelerated enormously by the defects. Therefore, an obvious question arises immediately as to how defects may cause so dramatic suppressing effect on ionic transport in mixed mobile ion systems?

This question can be answered by referring to the theory of ionic conductivity based on the defect diffusion model of Bendler and Shlesinger [64]. It predicts, in general formulation, that the rate of the diffusion <u>increases</u> with increase in number density of <u>single</u> defects, and <u>decreases</u> when for whatever reason defects are brought closer together and become <u>clustered</u> [65].

We argue here that in mixed alkali glasses, as well as in other mixed mobile ion systems, the way in which generation of structural defects occurs in the nearest vicinity of the guest-occupied sites, leads to the formation of local atomic configurations where mobile ions directly interact with more than one ion of the opposite sign, in mixed alkali glasses these are non-bridging oxyanions. These 'non-equilibrium' ionic sites with clustered anions are assumed to act as high-energy anion traps in glass network. (Note that equilibrium alkali sites also behave as anion traps of lower binding energy [66].)

The defect cluster can be characterized by its Coulombic potential and the capture cross section, $\pi d^2$, which are directly proportional to the conjugated negative charge of the cluster, $Z_o e$. The capture cross section defines the effective area of a trap, its capability to capture a charge carrier, and it is also a function of temperature $T$. The capture cross-section radius, $d$, is given by [66]

$$d = \frac{Z_o Z_i e^2}{8\pi \varepsilon_o \varepsilon \, k_B T} \tag{7}$$

where $\varepsilon_o$ and $\varepsilon$ are vacuum and relative dielectric constants respectively, $k_B$ is the Boltzmann constant, and $Z_i e$ is the charge of a charge carrier. The trapping event occurs when a mobile ion with the mass $m_i$ and mean thermal drift velocity $v_i = (2k_B T/m_i)^{1/2}$ approaches an anion trap at a distance less than $d$. Since the energy of an anion trap and its capture cross-section radius are directly proportional to the number of the clustered oxyanions, such a cluster with large $Z_o$ can act as a trapping site for more than one charge carrier attracting and capturing mobile ions from the neighboring regular ionic sites. In other words, the clustering of oxyanions and formation of high-energy anion traps can lead to the redistribution of the mobile ions in the vicinity of an anion trap, finally to the aggregation of the mobile ions. We will return to this point later in this section.

Now, with this picture in mind, we can evaluate how the changes in the local structure brought by high-energy anion traps transform the energetics of mobile ions in mixed ionic glass.

Random energy landscape in glass can be viewed as being composed of potential wells and structural barriers. In the model of Anderson and Stuart [34], the first term is the Coulombic interaction between an ion and its site, $\Delta E_B$, and the second one is an elastic strain energy, $\Delta E_S$, associated with the dilatation of the network forming matrix to allow an ion to pass from one site to another [34]:

$$\Delta E = \Delta E_B + \Delta E_S \qquad (7)$$

Let us assume for simplicity that a single-site anion trap is a spherical cavity comprised of oxygens of which only <u>two</u> are non-bridging oxyanions and the rest are bridging. This simplification is quite plausible in case of *Li-Na* or *Na-K* mixed alkali glasses. Although an interaction between alkali cation and its site is determined by the collective attractive potentials of their nearest environment composed of non-bridging and bridging oxygens, we assume here that bridging *Si-O* bonds are entirely covalent in character, so the electrical charge on the bridging oxygens is taken to be zero. According to Elliott [67], this simplification leads only to a small quantitative, but not qualitative change in the results. Like Isard [68] we assume also that the conjugate charge of two non-bridging oxyanions is delocalized over the entire spherical cavity, so that alkali cation with the charge $Z_i = +e$ resides in the cage uniformly carrying charge $Z_o = -2e$. This simplification enables us to treat the electrostatic interaction between an alkali cation and a single-site anion trap as pairwise. Then, since the formal charge of the site is doubled, the Coulombic site energy shall be doubled too. In the further simplification, the contribution of the strain energy term in activation energy, $\Delta E_S$, is taken to be the same both in single and mixed alkali glass. Based upon this simplified model, activation energies for the $Li^+$, $Na^+$, and $K^+$ cations in silicate glass in single-site anion traps have been calculated and the results are summarized in the Table III in comparison with those in equilibrium sites. Calculations were made using literature data on activation energy for ionic conductivity and calculated elastic strain energy in binary alkali silicate glasses. For the sake of definiteness, the number density of alkali cations in all glasses was taken as being ~$10^{22}$ cm$^{-3}$ which approximately corresponds to the trisilicate composition $xA_2O\text{-}(1\text{-}x)R_2O\text{-}3SiO_2$.

TABLE III. Activation energy, $\Delta E_E$ [84], strain energy, $\Delta E_S$ [34], calculated Coulomb binding energy in equilibrium sites, $\Delta E_B$, and activation energy in anion traps, $\Delta E_N$, for various alkali ions in mixed alkali silicate glasses.

|    | $\Delta E_E$, $eV$ | $\Delta E_S$, $eV$ | $\Delta E_B \cong \Delta E_E - \Delta E_S$, $eV$ | $\Delta E_N \cong 2\Delta E_B + \Delta E_S$, $eV$ |
|---|---|---|---|---|
| *Li* | 0.67 | 0 | 0.67 | 1.34 |
| *Na* | 0.72 | 0.10 | 0.62 | 1.34 |
| *K* | 0.75 | 0.35 | 0.42 | 1.15 |

Presumably, more realistic and precise calculation of the activation energy for the ions in single-site anion traps can be made based upon the method developed by Elliott where in addition to Coulombic forces, polarization and repulsion terms in the interaction potential as well as the geometry of the sites occupied by the cations and their concentration can be taken into account [67]. These seem to be important in case of multi-anion-cation configurations where charge screening effects might play a significant role.

The dc conductivity, $\sigma_{dc}$, is a product of the mobile charge density, $q = n\,Z_i e$, and the charge carrier microscopic mobility, $u$.

$$\sigma_{dc} = qu \equiv nZ_i eu \qquad (8)$$

where $n$ is the number density of the mobile ions and $Z_i e$ is the ion charge. The mobility can be defined, in turn, as a ratio of the mean (thermal) drift velocity in the direction of the applied electrical field to the magnitude of the applied field [66]. Strictly speaking, $u$ is also a function of $n$ but we neglect it here.

In the following, we will be mostly interested in the first quantity, the fraction of the charge carriers, since variation in the number density of mobile ions, but not their mobility, causes a strong conductivity dependence on the mole fraction of alkali oxide [69].

The number density of mobile ions is a temperature-dependent function, and it is given by the Boltzmann distribution

$$n(T) = n_O \exp(-\Delta E / k_B T) \tag{9}$$

where $n_o$ is the total ion number density and $\Delta E$ is the average activation energy.

Due to Coulombic and structural disorder in glasses, the activation energy for ion migration varies from site to site [70], so the equation (9) can be rewritten as [66]

$$n(T) = \sum_j n_j \exp(-\Delta E_j / k_B T) \tag{10}$$

where $n_j$ is the number density of ions residing in the sites with the activation energy $\Delta E_j$ and sum ranges over all ionic sites in the system.

For ions in binary alkali silicate glasses the variation in activation energy is well described by the single distribution function that is symmetric about a mean with well defined standard deviation. For mixed alkali glasses this is apparently not a case. Frischat [71] has studied self-diffusion of sodium in *Na-K*-silicate and aluminosilicate glasses, where potassium is a minority species, and concluded that the diffusion profile of sodium reflects a superposition of two diffusion processes with very different coefficients. This may only occur when there are (at least) two subsets of sodium sites in glass with very different activation energies for ionic migration.

One might expect that, likewise, two corresponding subsets of potassium sites should exist. Evidently, this is not a case. As is well known, the fraction of ions in process of hopping at any instant is extremely small [70]. Recent measurements indicate [72] that in soda silicate glass where total alkali content is $\sim 10^{22}$ $cm^{-3}$, the charge carrier number density does not exceed $(3-4) \cdot 10^{16}$ $cm^{-3}$. At room temperature, a sodium ion might sit in its regular site before moving on for approximately 1 second [73]. Nevertheless, we presume here that the vast majority of ions residing in equilibrium sites are mobile on the long timescale [74]. Consequently, on the long time scale, roughly all the minority ions in mixed alkali glass will be involved in site exchange with majority ions and convert all the minority equilibrium sites and a fraction of majority equilibrium sites into high energy anion traps. The compelling support to this conclusion can be found in the paper of Moynihan and co-workers who have studied electrical conductivity and relaxation in $xK_2O$-$(1-x)Na_2O$-$3SiO_2$ glasses and concluded that in the dilute sodium composition the $Na^+$ ions are essentially immobile, whereas a decrease in electrical conductivity may be considered as due to a decrease in the number of mobile $K^+$ ions [75]. A similar effect occurs at the other composition extreme with the roles of $Na^+$ and $K^+$ interchanged. A rapid decrease in electrical conductivity in very dilute composition and less rapid at larger foreign alkali content shall force us to the conclusion that the clustering of oxyanions leads to the aggregation of mobile ions in multi-anion-cation (MAC) clusters whose content and size are defined by the alkali composition and the size mismatch between dissimilar mobile ions. Presumably, an approach to the quantitative description of the content of the MAC clusters may be developed on the basis of study by Florian *et al.* [76] who investigated the cation distribution in mixed alkali disilicate glasses using 2D $^{17}O$ DAS NMR and demonstrated that evolution of the NBO isotropic line shape with changing alkali composition is consistent with the model of the glass structure where four alkali cations are distributed in random combinations around each NBO. In a glass of alkali composition $(Na_xK_{1-x})_2O$-$2SiO_2$, the combinations can be described by the binominal distribution according to which the number of NBO sites whose environment is $Si$-$O(Na_kK_{4-k})$ is given by:

$$P(k) = \frac{4!}{k!(4-k)!} x^k (1-x)^{4-k} \qquad (11)$$

where $k=0,...,4$.

Thus, without loss of generality within our simplified model, we may suggest that, depending on which alkali species is a minority, only two major subsets of alkali sites simultaneously coexist in the mixed alkali system: equilibrium single alkali sites $A_E$ or $R_E$, and MAC aggregates where both $A^+$ and $R^+$ are trapped. In the dilute foreign mobile ion composition, the aggregates are staffed largely with the majority species causing a drastic decrease in their number density residing in regular ionic sites. At the same time, activation energy for ions residing in regular ionic sites increases logarithmically with decrease in their number density [15]:

$$\Delta E_{A\ or\ R,E}\ (n_{A\ or\ R,E}) \propto\ ln(1/n_{A\ or\ R,E}) \qquad (12)$$

Thus, since nearly all minority species and a fraction of the majority species are largely immobilized in anion traps, mixed alkali glass behaves (at least when the concentration of the second alkali is not too large) as single alkali glass of effectively lower concentration of the charge carriers.

In the case of the symmetric situation when both unlike alkali ions have equal probability to enter host or foreign sites, on the long timescale nearly all the regular ionic sites will be converted into anion traps. It is important to emphasize that this occurs when the mobile fractions of both alkali species are equal and not at the composition where $n_A = n_R$.

One of the possible approaches to obtain analytical description of the compositional variation of the dc ionic conductivity of mixed alkali glass is to treat it as a randomly arranged binary mixture of two pseudophases, 'regular' and 'defective', with very different conductivities. It implies that all immobilized alkali species are part of the defective pseudophase, which mostly does not contribute to the ionic conduction until its volume fraction rises above a certain point. Beyond that point, the ions no longer can utilize purely the regular pseudophase pathways to move through the material. The volume fractions and conductivities of pseudophases are defined by the alkali composition. When the mobile fractions of both alkali species are equal, the regular pseudophase vanishes and ionic conduction occurs entirely within the defective pseudophase. At low concentrations of the second alkali, the conduction takes place entirely within the regular pseudophase, and it is governed by either $A^+$ or $R^+$ ions depending on which species is a majority. Since there are hard and easy pathways for ionic conduction through the material, the problem of ionic conductivity of such a 'composite' system can be solved within the framework of the continuum percolation model (effective medium approximation). The commonly used equation for the effective conductivity of two-phase composite system has been derived in its original form by Landauer for the conduction in two-component mixtures [77]. Later the theory was generalized for multi-phase composites by Wu and Liu [78]. The effective ionic conductivity of the two-phase composite is given by

$$\sigma_i = \frac{\varepsilon_1 + \sqrt{\varepsilon_1^2 + \varepsilon_2}}{4} \qquad (13)$$

where

$$\varepsilon_1 = 3(\upsilon_1\sigma_1 + \upsilon_2\sigma_2) - (\sigma_1 + \sigma_2)$$

$$\varepsilon_2 = 8\sigma_1\sigma_2$$

in which $\sigma_1$ and $\sigma_2$ are ionic conductivities of phases *1* and *2*, and $\upsilon_1$ and $\upsilon_2$ are their volume fractions, respectively. Since conductivities of the regular and defective pseudophases and their volume fractions change in nonsimple manner as the alkali composition of the mixed alkali system change, the application of the continuum percolation model would require the knowledge of the conductivity and volume fraction of each pseudophase for each alkali composition which are not currently available. We leave it for the future investigation. What we can do here is to estimate the decrease in conductivity of mixed alkali glass of symmetric alkali composition $n_A = n_R = n_o/2$ when only defective pheudophase exists. In this case, Eq. (10) reduces to two terms:

$$n(T) \cong \frac{n_O}{2}\{\exp(-\Delta E_{A,N}/kT) + \exp(-\Delta E_{R,N}/kT)\} \qquad (14)$$

where $\Delta E_{A,N}$ and $\Delta E_{R,N}$ are activation energies for $A^+$ and $R^+$ ions in single-site anion traps. For *Na$_2$O-K$_2$O-3SiO$_2$* glass at 500K the model predicts a decrease in charge carrier number density of order of $10^6$ with respect to that in binary silicate glass, which is in good agreement with literature data (see Table IV).

The other promising approach to obtain an analytical description of the compositional variation of the dc ionic conductivity of mixed alkali glass is to use a routine formulation of the percolation theory as proposed by Baranovskii and Cordes [20] where an unphysical assumption that cations $A^+$ cannot occupy sites $R$ and vice versa is replaced with a realistic description of the energy landscape were regular ionic sites and single-site anion traps coexist.

TABLE IV. Activation energy, charge carrier number densities in mixed alkali glass $xNa_2O\text{-}(1\text{-}x)K_2O\text{-}3SiO_2$ for $x = 0$; 0.5; and 1.

| $x$ | $\Delta E_{Na+}$, $eV$ | $n_{Na,E}$ | $n_{Na,N}$ | $\Delta E_{K+}$, $eV$ | $n_{K,E}$ | $n_{K,N}$ | $n_{total}$ |
|---|---|---|---|---|---|---|---|
| 0.0 |  | 0 | 0 | 0.75 | 2.7 $10^{14}$ | 0 | 2.7 $10^{14}$ |
| 0.5 | 1.34 | 0 | 1.5 $10^8$ | 1.15 | 0 | 5 $10^8$ | 6.5 $10^8$ |
| 1.0 | 0.72 | 5.4 $10^{14}$ | 0 | 1.15 | 0 | 0 | 5.5 $10^{14}$ |

## 5. MMIE IN VISCOSITY, $T_g$, AND MECHANICAL LOSS SPECTRA

In addition to ionic conductivity and diffusivity, glass viscosity and transformation temperature, $T_g$, which are not directly dependent upon ionic transport, show pronounced departure from linearity at intermediate mixed mobile ion composition. Reflecting the changes in glass structure, mechanical loss spectra also vary dramatically with addition of the second alkali. Within the framework of the proposed model, this behavior of the mixed mobile ion glasses receives simple and natural explanation.

In general, a transition from elastic solid to viscous fluid can be interpreted as depolymerization of the dynamic network-forming matrix. The mean size of the statistically polymerized regions becomes smaller with increasing temperature [79]. Therefore, the origin of the maximum in mechanical losses corresponding to glass transition is the rearrangement processes of structural units produced by opening and closing frequencies of bonding and diffusion processes of more or less large atomic complexes [80, 81]. Bond break-up and defect generation accompanying the rearrangement of ionic sites when ions jump into foreign sites in mixed mobile ion glasses increases the fragmentation of the glass network, which in turn results in the reduction of viscosity and in the shift of $T_g$ to lower temperatures. In the mechanical loss spectra, the shift indicates a loosening up of glass structure, which accompanies even small additions of the second mobile species. For the given concentrations of dissimilar mobile ions, this effect is the larger the greater the ion size mismatch.

Below $T_g$, much smaller characteristic maxima in mechanical loss spectra of relaxation type are observed. In general, these losses can be related to the stress-induced diffusional hops of mobile ions, and to the reorientation of such dangling structural moieties in glasses as non-bridging oxygen or stress-induced diffusion of oxygen ions or molecules. Studies of the mechanical losses of single alkali silicate glasses reveal an existence of two characteristic maxima in a fixed frequency (~ *1 Hz*) variable temperature scan [82, 83, 84, 85]. One maximum is usually observed at or below 0$^o$C, and the second appears between 100$^o$C and 300$^o$C. The magnitudes above background of the both maxima are almost linearly dependent on the alkali content [83, 86].

There is strong consensus in the literature that a maximum at low temperature is generally assigned to the stress-induced movement of alkali ions [82, 87]. This attribution is based upon the agreement between the activation energy calculated for this peak (*0.65 – 0.87 eV* [83], *0.65 – 0.69 eV* [84] for sodium), and that calculated for the dc electrical conductivity and alkali diffusion (*0.64 eV* [88]). Its progressive enlargement and movement to lower temperatures with increasing alkali content in single alkali glasses gives additional support to this conclusion [84].

The second maximum in mechanical losses has higher activation energy, and its structural origin has been ascribed to the relaxation processes associated with the non-bridging oxygen ions [87, 89, 90]. This peak becomes smaller and shifts to higher temperatures with changes in glass composition systematically eliminating NBOs [89, 90], and it becomes larger with increase in the alkali content, reflecting the fact that in ionic sites alkali cations are associated with NBOs. Although later Coenen [91], Makled and Kreidl [92], and Doremus [93] linked this maximum to the presence of water in glasses, and Day and others [94, 95, 96, 97] proposed that this maximum is caused by 'cooperative' motion of alkali ions and protons, similar to that suggested to explain the origin of so-called 'mixed alkali peak' in mechanical losses of mixed alkali glasses, this ascription seems to be incorrect since such a maximum is observed also in mechanical loss spectra of glasses whose IR spectra show them to be essentially anhydrous [73]. In addition, dielectric relaxation measurements which are usually used to qualify the response of the material to a field-induced perturbation revealed only maxima attributable to the motion of alkali ions and showed no response corresponding to the orientational polarization of dipoles proposed to be the centers responsible for this maximum [98, 99, 100, 101].

Measurements of mechanical losses of mixed alkali silicate glasses reveal the following typical changes. With addition of the second alkali to the glass composition, the low temperature alkali peak rapidly diminishes in amplitude and moves to higher temperatures. On the basis of the conductivity model outlined in the previous section, we can conclude that this behavior is reflecting the fact that fewer alkali sites respond to the applied stress due to immobilization of almost all minority alkali species and part of the majority species in high energy anion traps and activation energy increase for the alkali content remaining in the regular ionic sites. In a mixed alkali glass with equal number density of dissimilar alkali cations this peak is not observed since almost all alkali content is immobilized in high-energy anion traps.

The most profound change in the mechanical loss spectra of mixed alkali glasses is the development of so-called 'mixed alkali peak'. With increase in concentration of the second alkali, it rapidly increases in magnitude and moves to lower temperatures (see for example Fig. 2A in the Ref. 84). Similar behavior is observed in single alkali glasses with increase in residual water content where protons play role of the second mobile ion. Usually, only this maximum in mechanical losses is observed in the systems containing approximately equimolar concentrations of dissimilar mobile ions due to its large magnitude. The origin of this maximum in mechanical losses was a subject of many controversial interpretations. Initially, Rötger [102] and Jagdt [103] suggested that the mixing the alkali ions causes the original alkali maximum to enlarge and move to higher temperature. Steinkamp *et al*. [104] by making very small additions of the second alkali showed that this is not a case. Shelby and Day [84] acknowledged that in many cases this maximum looks like enlarged non-bridging oxygen one, however arrived to the conclusion that this was new 'mixed alkali peak' and related it to the cooperative motion of unlike alkali ions, specifically to the reorientation of elastic dipoles, which is controlled by the slower moving alkali ion. However, any motion of charge carriers, cooperative or not, shell be detectable by electrical field relaxation or dielectric relaxation methods whereas the process causing the 'mixed alkali peak' is electrically inactive, i.e. no dielectric loss peak was found that correlate with this maximum [85, 105]. Although it was suggested that the elastic dipoles might be electrically inactive, this attribution, like in case of high temperature maximum in spectra of single alkali glasses, should be considered incorrect.

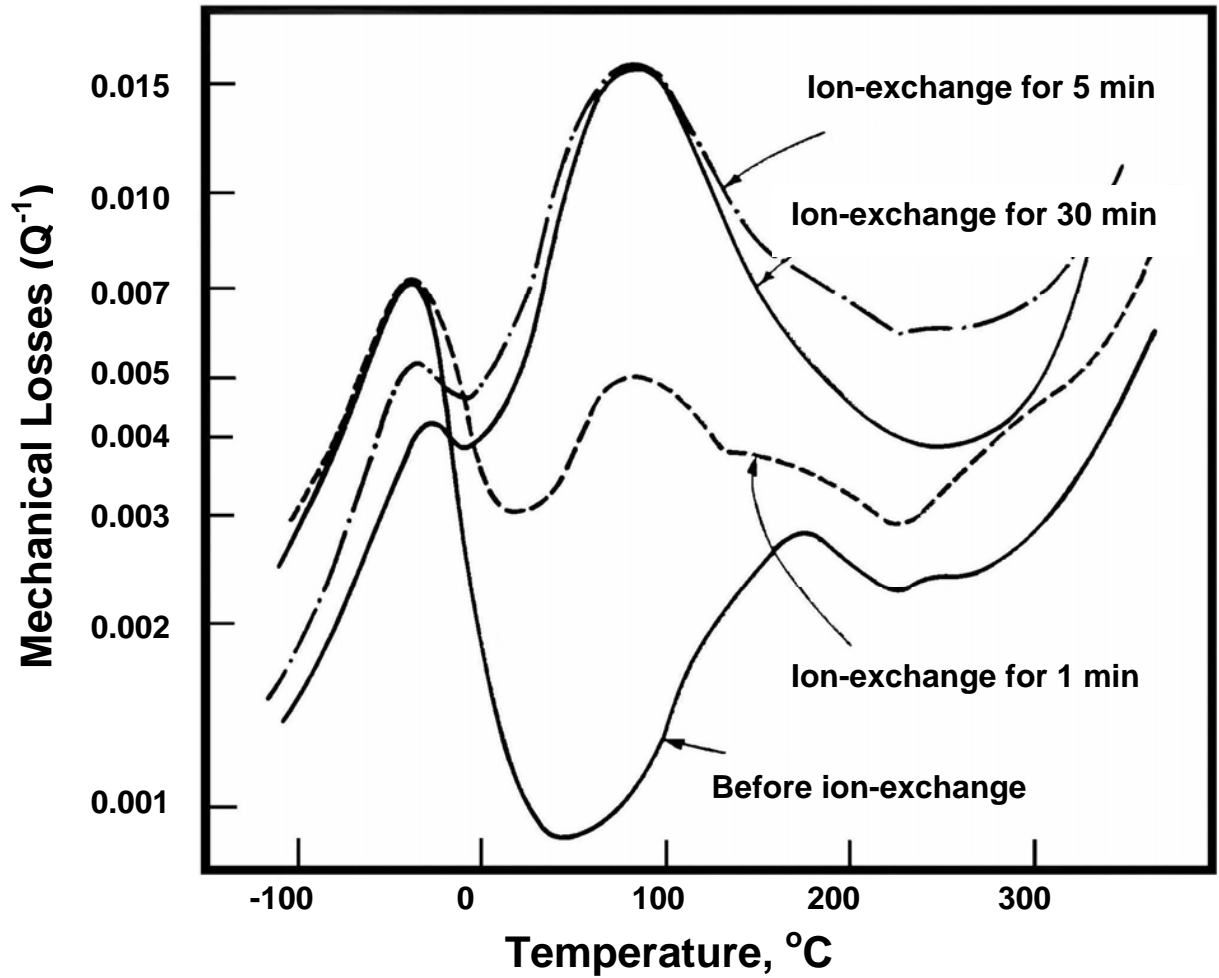

Fig. 2. Mechanical loss spectra of *Li$_2$O-2.5SiO$_2$* glass ion-exchanged in molten *NaNO$_3$* for various times (Redrawn from Ref. 122)

Comparison of the mechanical loss spectra of single alkali glass and mixed alkali glass with very small additions of the second alkali, as well as the fact that both high temperature loss maximum and 'mixed alkali peak' are electrically inactive, and employment of IR reflectance spectra and mechanical loss spectra of glasses subjected to ion exchange shown in Fig. 1 and Fig. 2 respectively as diagnostic make it immediately clear that so-called mixed alkali peak in mechanical losses is the enlarged in amplitude maximum assigned to non-bridging oxygen. The fact that with the change in relative concentrations of alkalis it may or may not shift to lower temperatures should not be confusing since glasses of different composition differ in density, elastic moduli, glass transition temperature etc. We should not be surprised also with the appearance of additional maxima in the mechanical losses of some mixed mobile ion glasses. Defect pair generation due to site rearrangement processes can give rise to various thermally activated reactions related to the annihilation of the oxygen vacancy centers which may occur by the trapping of diffusing molecular species such as $H_2O$, $O_2$, or $H_2$ [106] and formation of new structural moieties like peroxy radicals or peroxy linkages [36]. The diffusional processes and reactions linked to the annealing of excessive NBOs at higher temperatures also may cause the appearance of additional maxima in mechanical losses of mixed mobile ion glasses. Of course, these issues require further experimental and theoretical investigation.

## 6. EFFECT OF TEMPERATURE ON MMIE

Experimental investigations show that deep minima in isotherms of the dc conductivity, viscosity, and $T_{ag}$ of mixed alkali glasses and other mixed ionic conductors become progressively less pronounced with the temperature increase and tend to disappear at temperatures well above $T_{ag}$ [107, 108, and 109]. The model in question offers a consistent and natural way to account for the microscopic nature of this behavior.

Evidently, in oxide glasses the magnitude of the MMIE in all its features is determined by the steady state population of Knobs arisen from the rearrangement of ionic sites in mixed ionic systems which always exceeds the value corresponding to the condition of equilibrium (one NBO per ionic site). Consequently, depending on the temperature, thermal annealing of excessive Knobs shall take place converting in this way anion traps back into the regular ionic sites. This, in turn, releases the movement of the mobile ions. However, escaped ion may jump over again in a foreign regular ionic site and convert it into an anion trap repeating the damage – annealing cycles. Thus, two competing processes take place, so that the effect of temperature on the MMIE is defined by the interplay between kinetics of the structural defect generation and annihilation.

In general, because the movement of the mobile ions is decoupled from the remainder within a broad temperature range [110], the mechanism of the site rearrangement including generation of structural defects in mixed ionic conductors is largely temperature independent whereas the mechanism (or mechanisms) of their annihilation is temperature dependent. As was mentioned in the Section 3, the characteristic time scale describing a single defect generation time, $d_o$, is of order of the atomic vibrational period, $\sim 10^{-12}$ s. On the other hand, defect annealing time is a function of activation energy for the defect annealing, $Q$, and temperature, $T$, and it is given by [111]:

$$\tau_A(Q,T) = \frac{h}{k_B T} \exp(Q/k_B T) \tag{15}$$

where $h$ and $k_B$ are the Planck and Boltzmann constants, respectively.

Taking this formulation as a basis, we can draw a physical picture of the temperature effect on the MMIE.

Evidently, the MMIE is not observed in the low-temperature region below 200K [112]. Within that temperature range, the difference between activation energies of unlike mobile ions is so sufficient that in fact the only one most mobile species appears to be a charge carrier whereas another one is 'frozen out', so that ion transport occurs between the sites of the same type as in a single-ionic glass.

An increase to moderate temperatures 'defreezes' the second mobile species leaving frozen structural environment, although partial annihilation of the structural defects, specifically oxygen vacancies resulted from the rearrangement of ionic sites, may occur by either their occasional recombination with NBOs or by the trapping of various molecular species ($H_2$, $O_2$, $H_2O$). In this temperature range the annealing occurs via diffusional processes, although diffusion is sluggish. Thus, in the moderate temperature region the MMIE is the most pronounced.

The MMIE magnitude decreases when the rate of the annealing of structural defects increases with the further temperature increase, and this process becomes progressively more likely until the rates of the annealing of structural defects and their generation approach comparable values. Like in fused silica, below $T_g$ this process proceeds largely via diffusion-limited reactions with the spectra of activation energies, whereas above $T_g$ annealing of NBOs may occur also via thermally activated shedding and recoupling with different activation energy to restore the basic Si-O-Si linkages [113].

Let us assume for simplicity that the mean activation energy for the NBO annealing is an order of *0.5 eV* [106]. Of course, it is not quite exact, but it allows to make some order-of-magnitude estimates for the annealing rates, and to explore the trend in the MMIE with the temperature increase in wide temperature range. Calculations demonstrate that at *500K* the characteristic time of the defect annealing, $\tau_A$, is of the order of *$10^{-8}$ s*, which is four orders of magnitude slower than that of the defect generation one, $\tau_O$. Apparently, $\tau_A$ and $\tau_O$ shall reach comparable values well above $T_g$ which is consistent with experimentally observed behavior [114]. However, even though in that temperature range the minima in isotherms of the conductivity and viscosity become undetectable, the measurements of mechanical loss moduli in ternary silicate melts at GHz frequencies (Brillouin light scattering) reveal that the MMIE does not disappear, it becomes short-lived [115].

## 7. ANOMALOUS MIXED ALKALI EFFECT IN ION-EXCHANGED GLASSES

So-called 'anomalous mixed alkali effect' was discovered by Tomandl and Schaeffer [116] who studied electrical properties of soda-lime-silica glass subjected to $K^+$-for-$Na^+$ ion exchange, and found out that locally the ionic resistivity reaches its maximum value at the surface where $K^+$ ions fully replaced $Na^+$ ions and not at the intermediate $Na^+/K^+$ concentrations as should be expected from a 'crossover' in ion diffusivities.

Recently Ingram *et al.* [117] re-examined the ion transport properties of ion-exchanged glasses by a.c. impedance and confirmed the earlier result of Tomandl and Schaeffer.

Currently accepted view holds that this is high compressive stresses located at the surface which inhibit the migration of ions by restricting the amount of volume available [117, 118]. However measurements show that the local (resurface) resistivity increases by much larger factor than that theory predicts [117].

The explanation of this 'anomalous' mixed alkali effect on the basis of the model in question is obvious and even trivial: At the resurface where $Na^+$ ions are fully replaced by $K^+$ ions, in fact all the ionic sites where $K^+$ ions reside are former $Na$ sites converted into high energy anion traps. This explains why the maximum in resistivity is observed in the resurface and not in the lower surface layer where $Na^+$ and $K^+$ have approximately equal concentrations.

Simple estimate shows that in float glass studied by Ingram and co-workers [117] the activation energy for the ionic conductivity in the resurface layer where $Na^+$ is fully replaced by $K^+$ shall increase from *0.78 eV* to *1.32 eV* which is consistent with experimental data.

CONCLUSIONS

In this paper for the first time a physical concept is introduced postulating a formation of intrinsic structural defects in condensed materials as a response of network-forming matrix to the simultaneous migration of two or more dissimilar mobile ions of unequal size.

On the basis of this concept, the first microscopic model is developed that provides a comprehensive, adequate, consistent and generally applicable fundamental explanation for the mixed mobile ion effect in all its aspects and agrees, at least qualitatively, with all experimental facts.

Although this work has arisen from the exploration of alkali silicate glasses and much of the data discussed is on such systems, the approach to the MMIE developed here is applicable to other mixed ionic systems such as crystalline or even liquid ionic conductors.

Hopefully, this paper would stimulate further experimental and theoretical investigations on the physics of condensed materials in general and the mechanism of ionic transport in glassy, crystalline and liquid electrolytes in particular.

ACKNOWLEDGEMENTS

The author expresses thanks to Dr. I. Shinder and Dr. I. Talmy for the critical reading of the manuscript.